\documentclass[runningheads]{llncs}
\usepackage[T1]{fontenc}
\usepackage{caption}
\usepackage{subfigure}
\usepackage{algorithm,algorithmic}
\usepackage[misc]{ifsym} 
\usepackage{graphicx}
\usepackage{amssymb}
\begin{document}
\title{GTSD: Generative Text Steganography Based on Diffusion Model}
\titlerunning{Generative Text Steganography Based on Diffusion Model}
%
\author{Zhengxian Wu\orcidID{0000-0001-7957-0441} \and
Juan Wen$^{(\textrm{\Letter})}$\orcidID{0000-0002-4199-2988} \and
Yiming Xue\orcidID{0000-0001-6500-3868} \and
Ziwei Zhang\and
Yinghan Zhou}
\authorrunning{Z. Wu et al.}
%
\institute{Collage of Information and Electrical Engineering, China
Agricultural University, Beijing 100083, China\\
\email{wenjuan@cau.edu.cn}}
\maketitle              
\begin{abstract}
With the rapid development of deep learning, existing generative text steganography methods based on autoregressive models have achieved success. However, these autoregressive steganography approaches have certain limitations. Firstly, existing methods require encoding candidate words according to their output probability and generating each stego word one by one, which makes the generation process time-consuming. Secondly, encoding and selecting candidate words changes the sampling probabilities, resulting in poor imperceptibility of the stego text. Thirdly, existing methods have low robustness and cannot resist replacement attacks. To address these issues, we propose a generative text steganography method based on a diffusion model (GTSD), which improves generative speed, robustness, and imperceptibility while maintaining security. To be specific, a novel steganography scheme based on diffusion model is proposed to embed secret information through prompt mapping and batch mapping. The prompt mapping maps secret information into a conditional prompt to guide the pre-trained diffusion model generating batches of candidate sentences. The batch mapping selects stego text based on secret information from batches of candidate sentences. Extensive experiments show that the GTSD outperforms the SOTA method in terms of generative speed, robustness, and imperceptibility while maintaining comparable anti-steganalysis performance. Moreover, we verify that the GTSD has strong potential: embedding capacity is positively correlated with prompt capacity and model batch sizes while maintaining security.
\end{abstract}

\section{Introduction}
\begin{figure}[h]
    \centering
    \includegraphics[scale=0.2]{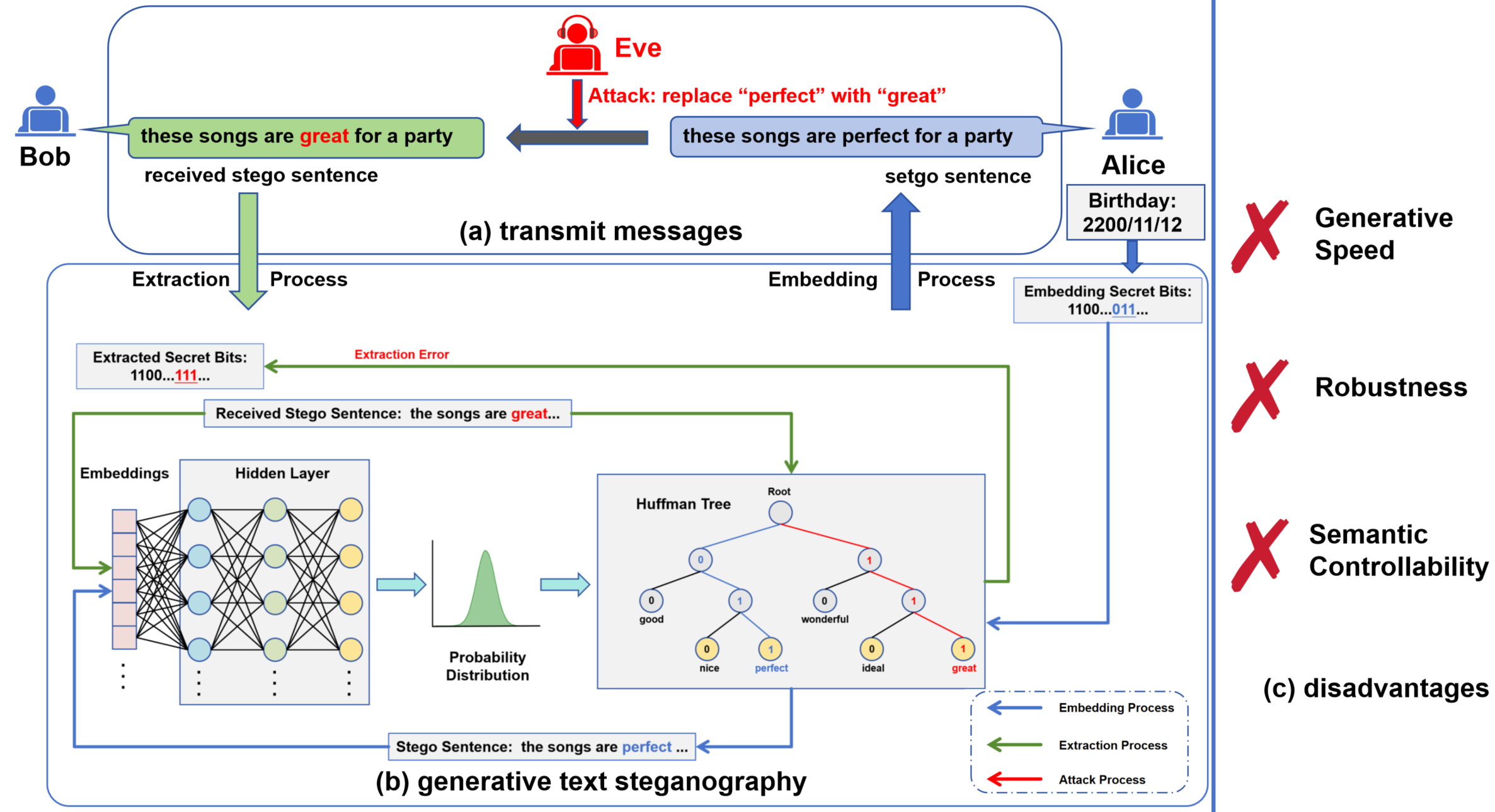}
    \caption{The process of existing generative text steganography methods.}
    \label{existing methods}
\end{figure}
With the continuous development of social networks, people are more and more inclined to communicate through images, audios, videos, text, etc. \cite{liao2020adaptive}, which greatly facilitates daily life but also increases the risk of privacy leakage. To safeguard data and communications, information security systems have emerged, with encryption systems and concealment systems being particularly prominent and receiving significant attention from researchers. The encryption system transforms readable plaintext into unreadable ciphertext to protect the security of secret information. Apart from that, the concealment system mainly protects secret information by hiding it into the common carrier, such as image \cite{liao2020adaptive}, audio \cite{wu2020audio}, video \cite{li2022anti}, and text \cite{zhou2021linguistic,zhang2021provably,ding2023discop}. This technology, also known as steganography, can conceal the existence of confidential information to achieve the purpose of not being easily suspected and detected. Due to the high accessibility, efficient transmission, and storage efficiency of text carriers, text steganography has attracted wide attention from researchers in recent years \cite{yang2020vae}. Text steganography can be divided into two categories: modification-based steganography \cite{mahato2013modified} and generative-based steganography. Modification-based text steganography embeds secret information by altering the given text carrier, such as format change\cite{ekodeck2016pdf}, synonym substitution\cite{Chen2008LinguisticSD}, and syntactic equivalent substitution\cite{DBLP}. However, since text has lower entropy and lacks redundancy for modification, these methods are low in embedding capacity and security.

In contrast, generative-based text steganography has become mainstream in recent years. Unlike modification-based text steganography, generative-based text steganography autoregressively generates innocuous-looking steganography straight from secret information using language models. As shown in Fig. \ref{existing methods}(a), assume the confidential message to be sent is "Birthday: 2200/11/12". Alice, the sender, first needs to turn the message into a bitstream "1100...011..." using any encryption algorithm. To conceal the existence of the bitstream, Alice uses text steganography to transform the bitstream into semantically innocuous but readable stego text, like "these songs are perfect for a party", and sends it to Bob under Eve's surveillance without arising Eve's suspicion. Let's further illustrate the main process of generative text steganography using Fig. \ref{existing methods}(b). Suppose the language model has previously output the stego phrase "the songs are". At this moment, "the songs are" is taken as input to obtain the probability distribution $p$ of the next token. Based on $p$, the top $n$ tokens with the highest probabilities are selected and Huffman coded. Suppose the bits to be embedded at this moment are "011". The token corresponding to "011", which is "perfect", is chosen as the stego output. Then "perfect" is appended to the end of "the songs are" and input into the model to obtain the probability distribution of the next token. This process repeats until all the bits have been embedded. To extract the confidential bitstream, Bob simply needs to repeat the above steps with the same start token and language model by comparing the received stego text. Generative-based text steganography methods have greatly alleviated the problem of low embedding capacity and security. However, it is prone to attacks, because extracted secrated bits are determined by received stego text. The most insidious and efficient attack is replacement attacks, such as random replacement, synonym replacement, and character substitution, which only need to change a few words to invalidate stego text. For example, Eve attacks the stego sentence by replacing "perfect" with "great". In this case, Bob extracts the wrong secret bits "111" which maps to "great". And, once an error occurs, the extraction of all subsequent bits will be impacted.

In this study, we argue that the autoregressive nature of language models causes cumulative errors in extracting secret bits when the stego message is attacked. Moreover, autoregressive language models generate tokens sequentially from left to right, leading to time-consuming and uncontrollable semantics. To address these issues, we abandon the classic autoregressive framework and adopt a non-autoregressive diffusion model, which is trained to learn the target distribution from a noise distribution. In recent years, diffusion models have been widely used in fields such as image generation and restoration due to their powerful generative capabilities. Recently, diffusion models have also been applied to text generation. Compared to autoregressive language models, diffusion models have three advantages: (1) Faster generation speed: Unlike autoregressive language models that generate tokens one by one, diffusion models predict all tokens simultaneously; (2) Strong redundancy: Diffusion models generate batches of text with the same semantics but different expressions; (3) High semantic controllability: The strong control capabilities of conditional diffusion models ensure that the generated text is highly controllable, and the generative priors of diffusion models guarantee the quality of the generated text. In this study, we utilize the AR-diffusion \cite{wu2024ar} as the basic framework to design text steganography method.

Inspired by the advantages of diffusion models, in this paper, we first introduce the diffusion model in text steganography and propose a new generative text steganography framework: Generative Text Steganography based on Diffusion model (GTSD). Our goal is to improve generative speed, robustness, and imperceptibility, while maintaining security. Specifically, we design a novel steganographic scheme based on diffusion model, mapping secret information to prompts and output batches. During the training process, a transformer-based denoising model denoises the noisy latent variable to learn the target probability distribution. During the inference phase, the prompts are used as controllable conditions to guide the pre-trained diffusion model to generate candidate texts in batches. Then, the stego text is selected from the batch of outputs. To reduce the significant number of inference steps, a skip mechanism is introduced to improve generative speed. In addition, since the diffusion model generates text in batches rather than token by token, using it for text steganography can significantly improve the speed of generating steganographic text. Extensive experiments demonstrate that the proposed method can achieve faster generative speed, stronger robustness, and imperceptibility while maintaining security. The contributions to this study are as follows:

$\bullet$ We propose a novel Generative Text Steganography based on Diffusion model (GTSD), marking the first use of non-autoregressive language models for generative text steganography. Compared to mainstream text steganography that generates stego tokens one by one, our model can efficiently generate all candidate stego texts at once.

$\bullet$ An embedding scheme is designed to embed sensitive information by selecting prompts and text from output batches. In addition, our method has strong potential: the larger the prompt capacity and model batch sizes, the larger the embedding capacity.

$\bullet$ We conduct extensive experiments on three classical text steganography datasets. All experiments demonstrate that our approach outperforms the SOTA text steganography methods in terms of generative speed, robustness, and imperceptibility, while maintaining comparable anti-steganalysis performance. Furthermore, experimental results verify that the embedding capacity of our method is positively correlated with prompt capacity and model batch sizes.

\section{Related Work}
With the development of Natural Language Processing (NLP) technology, generative text steganography has achieved remarkable success.

Le et al. \cite{van2003efficient} first applied the duality of steganography and source coding (i.e., data compression) to propose the generative text steganography based on arithmetic coding (AC). However, the AC-based method requires encrypting the message in advance and also relies on a more stringent condition: explicit data distribution. Yang et al. \cite{yang2018rnn} proposed a linguistic steganography based on recurrent neural networks that can automatically generate high-quality stego text on the basis of a secret bitstream that needs to be hidden. Zhou et al. \cite{zhou2021linguistic} proposed a linguistic steganographic model based on adaptive probability distribution and a generative adversarial network, which eliminates exposure bias and embedding deviation during the embedding process. To further improve security, Zhang et al. \cite{zhang2021provably} proposed a provably secure steganography method based on adaptive dynamic grouping (ADG). The ADG method dynamically grouped and numbered the probability distribution of all tokens in the vocabulary into groups with approximately the same probability sum and randomly sampled the next token from the normalized distribution of the group to which the secret bits correspond. The ADG-based generative text steganography achieves the SOTA method in terms of security. Although current generative text steganography methods have made significant progress and greatly improved anti-steganalysis performance, they still suffer from low robustness and slow generative speed.

\section{Generative Text Stegonagraphy based on Diffusion Model}

\subsection{Problem Definitions}
In this paper, we denote the secret bitstream to be hidden as $B=\{A_{1},A_{2},..., A_{n}\}$, where $A_{i}$ represents the bit segment read in the $ith$ time, consisting of a fixed number of binary digits. Its length affects the embedding capacity, which will be analyzed later. The generated stego text set is denoted as $S=\{s_{1},s{_{2}},...,s_{n}\}$, where $s_{i}$ is obtained by segment $A_{i}$ and the pre-trained diffusion model. After transmission over the public channel, the stego texts $S$ may be changed into $S'$ by a random replacement attack. Based on $S'$, the extracted bitstream, denoted as $B'$, is expected to be equal to $B$. In this paper, our goal is to ensure that our proposed steganography framework meets the general metric criteria for steganography \cite{4370824}: (1) \textbf{Security}: $S$ cannot be detected as abnormal by the mainstream steganalysis methods. (2) \textbf{Capacity}: The embedding capacity of the text steganography model, i.e., the number of bits embedded in one word, should be comparable to the SOTA steganography methods. (3) \textbf{Robustness}: When the stego text is subjected to random replacement attacks, the receiver can still accurately extract the secret information. In other words, $B=B'$. (4) \textbf{Imperceptibility}: The KL Divergence (KLD) between the statistical distributions of the stego texts and normal texts should be minimized as much as possible.
\begin{figure}[h]
    \centering    
    \includegraphics[scale=0.35]{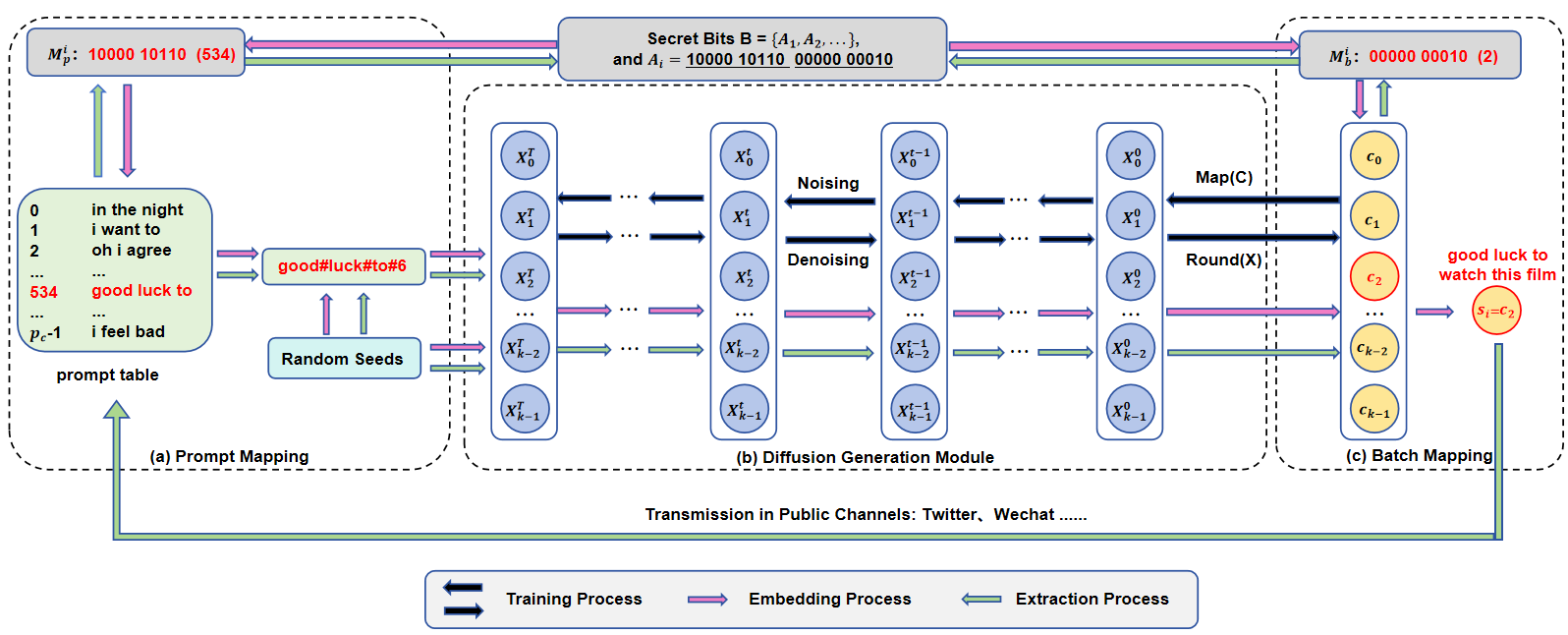}
    \caption{The architecture of proposed method. The $p_{c}$ and $k$ denote the capacity of prompt table and model batch sizes, respectively.}
    \label{structure}
\end{figure}
\subsection{Methodology}
The overall architecture of the proposed GTSD is shown in Fig. \ref{structure}, which consists of a prompt mapping module, a diffusion generation module, and a batch mapping module. 

\textbf{Diffusion Training.} \label{section3.2} 
Compared to images, which are composed of continuous pixels, texts consist of a discrete sequence of symbols (letters, words, etc.) with semantic associations, making them complex and abstract \cite{strudel2022self}. To project continuous text representations into the discrete feature space, text diffusion models generally utilizes embedding and rounding algorithms \cite{li2022diffusion}. As shown in Fig. \ref{structure} (black lines), GTSD conducts diffusion training in two processes: forward noising and backward denoising, with the backward denoising being guided by conditional prompts.

During the forward noising, the $j$th sentence $c_{j}$ of the sentence batch is mapped to a latent variable $x_{j}^{0}$, and gradually projected to a latent Gaussian distribution vector $x_{j}^{T}$ over $T$ time steps, by adding randomly sampled Gaussian noise $\epsilon\sim\mathcal{N}(0,I)$ in each step. $x_{j}^{t}$ represents the latent variable of sentence $c_{j}$ in time step $t$. The process of converting $x_{j}^{t-1}$ to $x_{j}^{t}$ is described as:
\begin{equation}
\label{equ1}
q(x_{j}^{t}|x_{j}^{t-1}) = \sqrt{\alpha_{t} } x_{j}^{t-1}+\sqrt{1-\alpha_{t}} \epsilon, t \sim [0,T],
\end{equation}
where $\alpha_{t}$ denotes a monotonically decreasing function from 1 to 0, and $T$ represents the total time length.

During the backward denoising, $x_{j}^{T}$ gradually removes the predicted noise $\epsilon_{\theta}$ to get the original embedding $x_{j}^{0}$, which can be rounded to sentence $c_{j}$. The process of converting $x_{j}^{t}$ to $x_{j}^{t-1}$ conditioned on prompt $P_{j}$ is described as:
\begin{equation}
\label{equ2}
p_{\theta}(x_{j}^{t-1}|x_{j}^{t};P_{j}) = \frac{x_{j}^{t}-\sqrt{1-\bar{\alpha} _{t}} \epsilon_{\theta}(x_{j}^{t},t,P_{j})}{\sqrt{1-\bar{\alpha} _{t}}},\\
\end{equation}
where $\bar{\alpha} _{t}= {\textstyle \prod_{n=1}^{t}\alpha _{n}}$, and $\epsilon_{\theta} \sim \mathcal{N}(0,I)$ is learned from the transformer model. $P_{j}$ denotes the conditional prompt of sentence $c_{j}$.

The forward noising and backward denoising can both be performed in batch training. The optimization of the diffusion model during the training can be described as follows: 
\begin{equation}
Loss  =  {\textstyle \sum_{t=0}^{T}} \mathbb{E} ||p_{\theta}(x_{0}|x_{t};P)-x_{0}||^{2}.
\end{equation}

\textbf{Mapping Mechanism.}
In generative text steganography, one of the most critical designs is establishing the mapping mechanism between the bitstream and the text. Traditional autoregressive steganography methods map the bitstream segments to word encoding, which is constructed by the output probability distribution of the language model at each time step. However, in diffusion models, we cannot access the probability distribution at each time step. Therefore, there is a need to design a novel mapping approach for GTSD.

In this study, we divide the bitstream $B$ into segments $\{A_{1},A_{2},... ,A_{n}\}$, each $A_{i}$ containing two parts, i.e., $A_{i}$=$(M_{p}^{i}, M_{b}^{i})$, where their lengths correspond to the size of the prompt table $p_{c}$ and model batch size $k$, respectively. Specifically, $M_{p}^{i}$ is used for selecting the specific prompt, which is used to generate a batch of candidates $C=\{c_1,c_2,...c_k\}$ based on backward denoising. The second block $M_{b}^{i}$ is used to select the corresponding stego text $s_{i}$ from $C$. For clarity, we refer to these two steps as prompt mapping and batch mapping, respectively.
 
 Prompt mapping necessitates a predefined prompt table. As shown in Fig. \ref{structure}, the prompt table consists of three words, which are selected from natural text. During prompt mapping, we map the bitstream block $M_{p}^{i}$ to a specific prompt in the prompt table. For example, when $M_{p}^{i}$ = "10000 10110", we select the No. 534 (convert binary to decimal) prompt "good luck to" and combine it with a randomly sampled length $l^{i}$, as the conditional prompt $P$ to guide the diffusion model to generate the candidate set $C$. Notice that the prompt of these three words cannot change during the generation process. To ensure this, we performed prompt filtration using a pre-trained diffusion model beforehand. 
 
After generating a batch of candidate sentences $C$ according to the conditional prompt $P$, batch mapping is applied to obtain the final stego text $s_{i}$ according to $M_{b}^{i}$. For example, when $M_{b}^{i}$ = "00000 00010", we select $c_2$ from $C$, ending up with the final stego text $s_{i}$ = "good luck to watch this film".

\begin{algorithm}[h]
\renewcommand{\algorithmicrequire}{\textbf{Input:}}
\renewcommand{\algorithmicensure}{\textbf{Output:}}
\caption{Information Hiding Algorithm}
\begin{algorithmic}[1]
\label{Algorithm1}
\REQUIRE $ $\\
secret bits $B$, prompt table $PT$ with a length of $p_{c}$, the pre-trained diffusion model, random seeds $r$, batch sizes $k$, and text length $l\in$ [$l_{min}$,$l_{max}$]
\ENSURE $ $\\
stego texts $S=\{s_{1}, s_{2}, ...\}$
\STATE  Sample initialized Gaussian distributions $N$=$\{N_{0},N_{1},...,N_{k}\}$ by random seeds $r$
\STATE  Divide secret bits $B$ into segments $\{A_{1},A_{2},...\}$ where each $A_i$ has a length of $p_{c}+k$
\STATE  Random sample length $l$=$\{l^{0},l^{1},...,l^{|A|-1}\}$ form $l_{min}$ to $l_{max}$ by random seeds $r$
\STATE  Set stego texts $S$=$\{$$\}$
\WHILE {not end of segments $\{A_{1},A_{2},...\}$}
\STATE  Divide $A_{i}$ into $M_{p}^{i}$ and $M_{b}^{i}$, with their lengths being $p_{c}$ and $k$, respectively
\STATE Extract the specific prompt corresponding to the decimal value of $M_{p}^{i}$, and combine it with $l^{i}$ to obtain the conditional prompt $P_{i}$
\STATE  Input $P_{i}$ to the diffusion model, get a batch of candidate sentences $C$=$\{c_{0},c_{1},...,c_{k}\}$ from Gaussian distributions $N$
\STATE  select stego text $s_{i}$ from $C$ according to $M_{b}^{i}$ by batch mapping
\STATE  $S$.append($s_{i}$)
\ENDWHILE
\RETURN stego texts $S$
\end{algorithmic}
\end{algorithm}

\textbf{Information Hiding Algorithm.}
 The pink lines shown in Fig. \ref{structure} present the overall information hiding process, which is occurred only during inference time and is described in Algorithm \ref{Algorithm1}. The core idea is using the pre-trained diffusion model to generate stego texts from randomly sampled Gaussian distributions $N$ under the guidance of the conditional prompt. Note that the initialized Gaussian distributions $N$ and length $l$ are generated by random seeds $r$. In this paper, the $l_{min}$ and $l_{max}$ are set to 5 and 25. To improve generative speed, we introduce a skip mechanism \cite{li2022diffusion,gong2022diffuseq} to decrease the number of sample steps in the inference stage. 
 
\textbf{Information Extraction Algorithm.} \label{section3.3}
The green lines in Fig. \ref{structure} present the information extraction process. The sender and the receiver share the same prompt table, pre-trained diffusion model, model batch sizes, and random seeds. The overall extraction process is described in Algorithm \ref{Algorithm2}.

\begin{algorithm}[h]
\renewcommand{\algorithmicrequire}{\textbf{Input:}}
\renewcommand{\algorithmicensure}{\textbf{Output:}}
\caption{Information Extraction Algorithm}
\label{Algorithm2}
\begin{algorithmic}[1]
\REQUIRE $ $\\
stego texts $S$ with $|S|$ sentences, prompt table $PT$ with a length of $p_{c}$, random seeds $r$, batch sizes $k$, and text length $l \in$ [$l_{min}$,$l_{max}$]
\ENSURE $ $\\
secret bits $B$ \\
\STATE  Sample initialized Gaussian distributions $N$=$\{N_{0},N_{1},...,N_{k}\}$ by random seeds $r$
\STATE  Random sample length $l$=$\{l^{0},l^{1},...,l^{|S|-1}\}$ range form $l_{min}$ to $l_{max}$ by random seeds $r$
\STATE  Set secret bits $B$=$\{$$\}$
\WHILE {not end of stego texts $S=\{s_{1},s_{2},...\}$}
\STATE  Match the first three words of $s_{i}$ and $PT$ to obtain the specific prompt and $M_{p}^{i}$
\STATE  Combine the special prompt and $l^{i}$ as conditional prompt $P_{i}$ to guide the pre-trained diffusion model generating a batch of candidate sentences $C$=$\{c_{0},c_{1},...,c_{k}\}$ from Gaussian distributions $N$
\STATE  Calculate the similarity score $Q$ of $s_{i}$ and each sentence $c_i \in C$
\STATE  Find $c_{*}$, which maps to the largest score $Q_{*}$, to obtain $M_{b}^{i}$
\STATE  $B$.append($A_{i}$=($M_{p}^{i}$,$M_{b}^{i}$))
\ENDWHILE
\RETURN secret bits $B$
\end{algorithmic}
\end{algorithm}

The receiver uses the shared random seeds $r$ to sample the initialized Gaussian distributions $N$ and length $l$. Then $M_{p}^{i}$ and the special prompt is obtained by matching the first three words of the received stego text $s_{i}$ based on the shared prompt table $PT$. After that, the receiver uses the same method as the sender to generate a batch of candidate sentences $C$ from Gaussian distributions $N$. Next, the receiver compute the similarity score $Q$ between $s_i$ and each sentence in $C$, and find the best score to get the $M_{b}^{i}$. Finally, the $ith$ segment of secret bits is obtained by combining $M_{b}^{i}$ and $M_{p}^{i}$. 

It is worth mentioning that even when the transmitted stego texts undergo replacement attacks, which change a few words that alter the stego texts, our method maintains a high accuracy in correctly extracting the secret information. This hypothesis is validated in Section \ref{4}.

 \textbf{Discussion on embedding Capacity.} The size of the prompt table and the model batch are positively correlated with the embedding capacity of GTSD.  The embedding capacity $bwp$ (bits per word) can be calculated as:
 \begin{equation}
 bpw=\frac{ \log_{2}{k}+\log_{2}{p_{c}} }{\bar{l} }
 \end{equation}
 where $\bar{l}$ is the average length of stego texts $S$. In Sec. \ref{4}, we varied the sizes of the prompt table and model batch sizes to get different embedding capacities for comparative analysis.
 
\section{Experimental Results and Analysis} \label{4}
\subsection{Setup}
To verify the performance of the proposed method, we conduct our experiments on three publicly available standard datasets: Movie \cite{maas-etal-2011-learning}, Twitter \cite{go2009twitter}, and News \cite{zhang2021provably}. Since the sentence embedding length of the diffusion model is 54, we filter sentences with sentence lengths less than 5 or greater than 54. Finally, we randomly divide the dataset into a training set, a validation set, and a test set according to the ratio of 8:1:1.

In this paper, we conduct experiments on NVIDIA RTX 4060 and CUDA 11.6. In the training stage, the batch sizes and initial learning rate are set to 128 and 1e-3, respectively. We utilize the AdamW as the optimization and dropout mechanism to prevent overfitting. During inference, the model batch sizes $k$ is set to 1024. Moreover, we design a prompt table and a extended prompt table, and their lengths $p_{c}$ are set to 1024 and 1024$^3$, respectively.
\subsection{Baselines}
We reproduce the following text steganography methods and steganalysis tools to verify the performance of the GTSD in terms of security and efficiency.

\textbf{Steganography methods.} We utilize VLC\cite{yang2018rnn}, AC\cite{van2003efficient}, and ADG\cite{zhang2021provably} as our baselines. Each steganography method generates 11,000 stegos and 11,000 covers for evaluating performance.

\textbf{Steganalysis methods.} We employ linguistic steganalysis methods, including BERT-F \cite{peng2021real}, LS-CNN \cite{wen2019convolutional}, TS-BiRNN \cite{yang2019ts}, and BiLSTM-Dense \cite{yang2020linguistic}, to test the anti-steganalysis ability of GTSD.  In our experiments, 'stego' is regarded as a positive class, while 'cover' is regarded as a negative class, indicating the absence of embedded secret information.

\subsection{Metrics}
\textbf{Embedding Capacity.} Embedding capability represents the average amount of bits that one single token can carry. This paper uses bits per word (bpw) as the measure, with higher values indicating greater capacity.

\textbf{Security.} Steganalysis tools mainly distinguish detected text as cover or stego. To evaluate the anti-steganalysis ability, accuracy and F1 are utilized as measurements, which are calculated as $Acc = \frac{TP+TN}{TP+FP+FN+TN}$, and $F1 = \frac{2TP}{2TP+FP+FN}$, where $TP$ and $TN$ are true positives and true negatives, $FP$ and $FN$ are false positives and false negatives.

\textbf{Imperceptibility.} We calculate the KL divergence (KLD) between the statistical distributions of the sentence embedding of cover and stego to indirectly reflect the overall information-theoretic security. The KLD is computed by:
\begin{equation}
D_{KL}(p_{c}||p_{s}) \approx \sum (log\frac{\sigma_{s}}{\sigma_{c}}+\frac{\sigma_{c}^{2}+(\mu_{c}-\mu_{s})^{2}}{2\sigma_{s}}-\frac{1}{2}),
\end{equation}
where $\mu_{c}$ and $\mu_{s}$ are the mean deviation of cover and stego vectors, while $\sigma_{c}$ and $\sigma_{s}$ are the standard deviation of cover and stego vectors. In this paper, the dimension of sentence vectors is set to 100.

\textbf{Robustness.} We attack the stego text by randomly replacing $n$ words in each sentence and use the Sentence Correct Extraction Rate (CER) after replacement attacks as a measure of robustness. That is to say, if any hidden bit in a stego sentence is not extracted correctly, the sentence is considered to have been extracted incorrectly. To avoid randomness, we perform 10 rounds of replacement attacks on 1,000 stego texts and calculate the Average Sentence Correct Extraction Rate (ACER):
\begin{equation}
ACER = {\textstyle \sum_{i=1}^{10}} \frac{CN_{i}}{1000 \times 10}
\end{equation}
where $CN_{i}$ denotes the number of correctly extracted stego sentences in the $i$th round. In this paper, $n$ is set to 0, 1, 2, 3, and 4.

\textbf{Generative speed.} To test the model efficiency, We measure the total time spent generating 512 stego texts and calculated the average time per stego text in seconds.

\subsection{Compared with Baselines}
We compare the proposed model with baselines in terms of security and embedding capacity, robustness, imperceptibility, and generative speed.

\begin{table*}[t]
\centering
\caption{The detection accuracy and F1 of stego texts. The -b (-e) mean using the prompt table (extended prompt table) during prompt mapping.}
\label{BPW}
\begin{tabular}{c|c|c|c|c|c|c|c|c|c|c}
\hline
Datasets & Models & bpw$\uparrow$ & \multicolumn{2}{c|}{BERT-F}&\multicolumn{2}{c|}{LS-CNN}&\multicolumn{2}{c|}{TS-BiRNN}&\multicolumn{2}{c}{BiLSTM-Dense}\\
\cline{4-11}
 & & & acc$\uparrow$ & F1$\downarrow$ & acc$\uparrow$ & F1$\downarrow$ & acc$\uparrow$ & F1$\downarrow$ & acc$\uparrow$ & F1$\downarrow$ \\
\hline
Movie & VLC & 1.82 & 0.9665 & 0.9665 & 0.9580 & 0.9580 & 0.9400 & 0.9400 & 0.9410 & 0.9410 \\
\cline{3-11}
 & & 4.41 & 0.9230 & 0.9230 & 0.8590 & 0.8590 & 0.8530 & 0.8530 & 0.8565 & 0.8565 \\
\cline{2-11}
 & AC & 2.21 & 0.9690 & 0.9690 & 0.9535 & 0.9535 & 0.9365 & 0.9365 & 0.9350 & 0.9350 \\
\cline{3-11}
 & & 3.90 & 0.9320 & 0.9320 & 0.8815 & 0.8815 & 0.8735 & 0.8735 & 0.8645 & 0.8645 \\
\cline{2-11}
 & ADG & \textbf{5.07} & 0.5184 & 0.4577 & 0.5055 & \textbf{0.4468} & 0.4985 & \textbf{0.3317} & 0.4970 & \textbf{0.3396} \\
\cline{2-11}
 & Ours-b & 1.87 & \textbf{0.5000} & 0.3430 & \textbf{0.4835} & 0.4741 & 0.5085 & 0.5079 & \textbf{0.4915} & 0.4808 \\
\cline{2-11}
 &  Ours-e & 4.97 & \textbf{0.5000} & \textbf{0.3333} & 0.5045 & 0.4751 & \textbf{0.4820} & 0.4803 & 0.4935 & 0.4305 \\
\cline{1-11}

Twitter & VLC & 1.81 & 0.9370 & 0.9370 & 0.9280 & 0.9280 & 0.8950 & 0.8950 & 0.8810 & 0.8810 \\
\cline{3-11}
 & & 4.48 & 0.8680 & 0.8679 & 0.8120 & 0.8120 & 0.8035 & 0.8032 & 0.7770 & 0.7769 \\
\cline{2-11}
 & AC & 2.13 & 0.9225 & 0.9225 & 0.9040 & 0.9040 & 0.8770 & 0.8770 & 0.8840 & 0.8840 \\
\cline{3-11}
 & & 3.93 & 0.8845 & 0.8845 & 0.8310 & 0.8310 & 0.8285 & 0.8284 & 0.8210 & 0.8210 \\
\cline{2-11}
 & ADG & 5.11 & \textbf{0.4995} & 0.3375 & 0.5145 & 0.5137 & 0.4965 & 0.3595 & \textbf{0.5000} & \textbf{0.3386} \\
\cline{2-11}
 & Ours-b & 2.20 & 0.5125 & 0.5110 & 0.4980 & 0.4659 & \textbf{0.4900} & \textbf{0.4830} & 0.5035 & 0.4963 \\
\cline{2-11}
 &Ours-e  & \textbf{5.37} & 0.5000 & \textbf{0.3333} & \textbf{0.4970} & \textbf{0.3789} & 0.5035 & 0.4418 & 0.5040 & 0.4049 \\
\cline{1-11}
 
 News & VLC & 1.82 & 0.9770 & 0.9770 & 0.9715 & 0.9715 & 0.9625 & 0.9625 & 0.9640 & 0.9640 \\
\cline{3-11}
 & & 3.91 & 0.9620 & 0.9620 & 0.9340 & 0.9340 & 0.9060 & 0.9060 & 0.9195 & 0.9195 \\
\cline{2-11}
 & AC & 2.17 & 0.9690 & 0.9690 & 0.9645 & 0.9645 & 0.9495 & 0.9495 & 0.9580 & 0.9580 \\
\cline{3-11}
 & & 3.99 & 0.9615 & 0.9615 & 0.9205 & 0.9205 & 0.9029 & 0.9029 & 0.8995 & 0.8994 \\
\cline{2-11}
 & ADG & 5.43 & 0.5250 & 0.5234 & 0.4925 & 0.4915 & \textbf{0.5005} & \textbf{0.3344} & 0.5005 & \textbf{0.338} \\
\cline{2-11}
 & Ours-b & 2.22 & \textbf{0.4970} & \textbf{0.3504} & \textbf{0.4230} & \textbf{0.3798} & 0.5235 & 0.5213 & \textbf{0.4170} & 0.3693 \\
\cline{2-11}
 & Ours-e & \textbf{5.49} & 0.5130 & 0.5893 & 0.5025 & 0.4937 & 0.5170 & 0.5148 & 0.5050 & 0.4786 \\
\cline{2-11}
\hline
\end{tabular}
\label{table_MAP}
\end{table*}

\textbf{Security and Embedding Capacity.}
 For a steganographic model, lower steganalysis performance indicates that it is more difficult to distinguish between cover and stego texts, which means the model's security is better. To verify the security of the proposed method, four steganalysis tools are used to detect stego texts, and the results are shown in Table. \ref{BPW}.From Table. \ref{BPW}, it can be seen that VC- and VLC-based steganography methods have relatively weak resistance against current mainstream steganalysis algorithms, with detection accuracy and F1 scores generally above 90\%. In contrast, the state-of-the-art steganography method ADG, based on provably secure design, shows anti-detection performance against various steganalysis tools consistently below 50\%. Our proposed GTSD demonstrates comparable steganalysis resistance to ADG. We also note that the embedding capacity of our method increases with the size of the prompt table. Although the embedding capacity of our method equipped with the basic prompt table is lower than ADG, GTSD achieves a higher embedding capacity with the extended prompt table. Unlike the ADG algorithm, which cannot control capacity (with bpw typically around 5), our algorithm can increase the hidden capacity by expanding the size of the prompt table.

\textbf{Robustness Analysis.}
Generative text steganography methods based on autoregressive language models lack robustness under replacement attacks. On the one hand, the VLC-based and AC-based methods use Huffman coding and arithmetic coding, respectively, for secret embedding. Since each word coding is unique, changing any word in stego text will lead to the fail of bitstream extraction, If every sentence is subjected to an attack, the proportion of correctly extracted sentences would be $0$. On the other hand, the ADG-based method embeds secret information based on dynamic groupings, which may contain multiple words in each group. If we assume that the replaced word $d_{i}$ belongs to group $D=\{d_{1},d_{2},...\}$ and the random replacement word $e_{j}$ belongs to candidate pool $E=\{e_{1},e_{2},...\}$. Then, the Sentence Correct Extraction Rate of ADG can be simply calculated as $\frac{|E\cap D|}{|E|}$. According to the principle of ADG, the vocabulary size is $50178$, the ideal average group number is $2^{5}$, and the average words in each group are $1568$, and $E$ is set to vocabulary. The theoretical Sentence Correct Extraction Rate of ADG is close to $\frac{1}{32}$ after replacing one word.

Compared with the existing generative text steganography methods, our proposed method has better robustness. Under non-attack conditions, we conduct extraction experiments with a successful extraction accuracy of 100\%. Then We randomly replace the words of each stego text $n$ times (except prompts) and verify the robustness with the ACER, as shown in Table. \ref{robustness}. When $n$=1, 2, and 3, the ACEA can basically reach about 95$\%$. Moreover, when $n$ = 4, the ACEA has decreased but still remains above 80$\%$, which is because our sentence length is set to short. As the batch size increases, there is a slight decrease in robustness, but the overall effect is small.
\begin{table}[t]
\centering
\caption{The Average Correct Extraction Rate (ACER) of the proposed GTSD. The left (right) of "$|$" means the attacked words without (with) prompts.}
\label{robustness}
\begin{tabular}{c|cc|cc|cc}
\hline
n        & \multicolumn{2}{c|}{1}                         & \multicolumn{2}{c|}{2}                         & \multicolumn{2}{c}{3}                          \\ \hline
batch sizes & \multicolumn{1}{c|}{128}         & 256         & \multicolumn{1}{c|}{128}         & 256         & \multicolumn{1}{c|}{128}         & 256         \\ \hline
News     & \multicolumn{1}{c|}{99.94$|$69.57} & 99.97$|$67.21 & \multicolumn{1}{c|}{99.29$|$47.43} & 99.41$|$47.08 & \multicolumn{1}{c|}{96.15$|$29.30} & 96.24$|$30.96 \\ \hline
Movie    & \multicolumn{1}{c|}{99.90$|$67.38}  & 99.91$|$70.59 & \multicolumn{1}{c|}{98.43$|$44.07} & 98.01$|$44.14 & \multicolumn{1}{c|}{93.29$|$26.95} & 91.04$|$25.70  \\ \hline
Twitter  & \multicolumn{1}{c|}{99.97$|$64.63} & 99.99$|$65.84 & \multicolumn{1}{c|}{99.71$|$41.16} & 99.58$|$41.18 & \multicolumn{1}{c|}{98.19$|$25.32} & 98.59$|$25.31 \\ \hline
\end{tabular}
\end{table}

\textbf{Imperceptibility.} We use KLD to measure the distortion between stego texts and cover texts, and the results of KLD are listed in Table. \ref{KLD}. For AC-base and VLC-based methods, we apply two different embedding payloads. As shown in Table. \ref{KLD}, the KLD decreases with the increase of bpw for AC-based and VLC-based methods. This is a common phenomenon in text steganography \cite{zhang2021provably}: as bpw increases, the candidate pool of stego text grows closer to the candidate pool of cover text, making the distortion decrease. Our method also basically satisfies this phenomenon, and with the increase of bpw, KLD has a certain increase. Compared to the ADG-based method, the GTSD has stronger imperceptibility, especially in Movie and News. In addition, the bpw of the ADG-based method cannot be increased, which means that the KLD cannot be reduced. But the GTSD can increase the bpw by improving the capacity of prompt table and model batch sizes.

\begin{table*}[hp]
\centering
\caption{The KLD of different steganography methods.}
\label{KLD}
\begin{tabular}{c|c|c|c|c|c|c}
\hline
Models & \multicolumn{2}{c|}{Movie}&\multicolumn{2}{c|}{Twitter}&\multicolumn{2}{c}{News} \\
\cline{2-7}
 & bpw$\uparrow$ & KLD$\downarrow$ & bpw$\uparrow$ & KLD$\downarrow$ & bpw$\uparrow$ & KLD$\downarrow$ \\
\hline
 VLC & 1.82 & 3.55 & 1.82 & 1.32 & 1.81 & 1.33 \\
\cline{2-7}
 & 4.41 & 0.74 & 3.91 & 3.40 & 4.48 & 2.26 \\
\cline{1-7}
AC & 2.21 & 2.71 & 2.17 & 1.43 & 2.13 & 3.91 \\
\cline{2-7}
 & 3.90 & 0.24 & 3.99 & 3.45 & 3.93 & 2.49 \\
\cline{1-7}
ADG & \textbf{5.07} & 0.32 & 5.43 & \textbf{0.35} & 5.11 & 0.16 \\
\cline{1-7}
Ours-b& 1.87 & 0.47 & 2.22 & 0.56 & 2.20 & \textbf{0.11} \\
\cline{1-7}
Ours-e& 4.97 & \textbf{0.08} & \textbf{5.49} & 0.54 & \textbf{5.37} & 0.12 \\
\hline
\end{tabular}
\label{table_MAP}
\end{table*}

\textbf{Generative Speed.}
Since diffusion models generate all tokens simultaneously, the generative speed of the proposed method is significantly faster than that of the baselines, as shown in Table \ref{TIME}. Compared to the baselines, the generative speed of GTSD is approximately 0.03 seconds per sentence, making it 50 to 100 times faster. This observation indicates that the proposed GTSD is well-suited for real-time covert communication.

\begin{table*}
\centering
\caption{The comparison of generative speed.}
\label{TIME}
\begin{tabular}{c|c|c|c}
\hline
Models & Movie & Twitter & News \\
\hline
 VLC & 3.526 & 1.511 & 3.762 \\
\cline{1-4}
AC & 0.336 & 0.185 & 0.417 \\
\cline{1-4}
ADG & 1.499 & 0.618 & 2.792 \\
\cline{1-4}
Ours-e & \textbf{0.034} & \textbf{0.036} & \textbf{0.039} \\
\hline
\end{tabular}
\label{table_MAP}
\end{table*}

\subsection{Ablation}
\begin{figure}
\centering  
\subfigure[bpw]{
\label{bpw}
\includegraphics[scale=0.3]{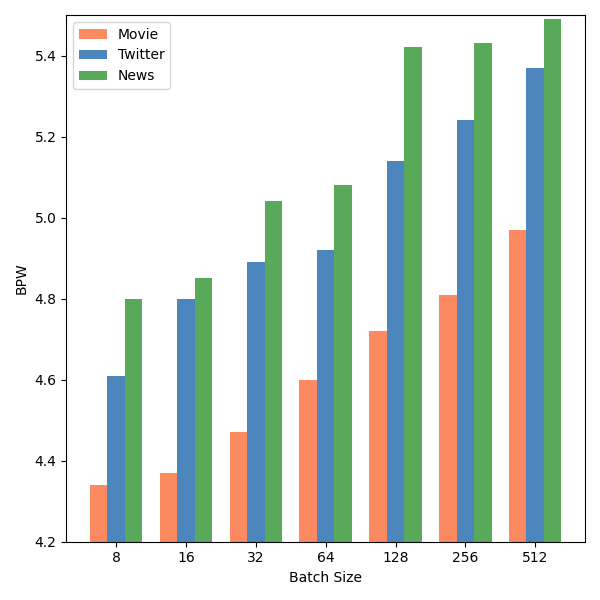}}
\subfigure[times]{
\label{times}
\includegraphics[scale=0.3]{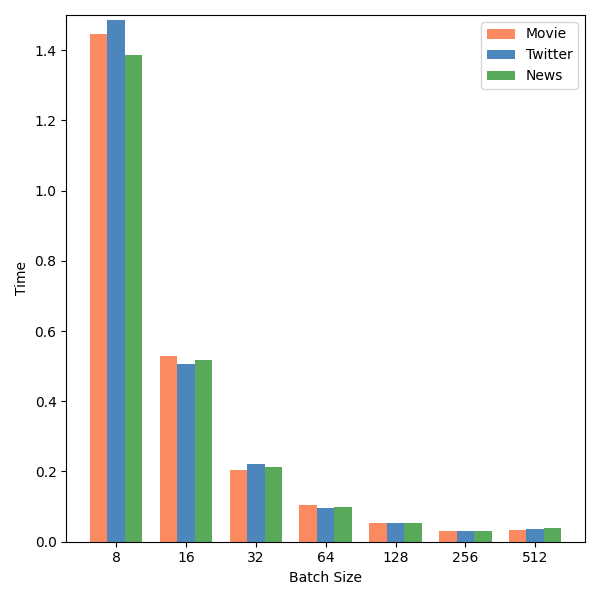}}
\caption{The bpw and times of different batch sizes.}
\label{results}
\end{figure}

To explore the potential of the GTSD, we design two ablation schemes: (1)We use basic prompt table and extended prompt tables during prompt mapping to observe the effects of prompt capacity on embedding capacity, anti-steganalysis ability, and imperceptibility. The results as shown in Table. \ref{BPW} and Table. \ref{KLD}; and (2) we set different model batch sizes during the batch mapping, including 8, 16, 32, 64, 128, 256, and 512, to observe the relationship between embedding capacity, generative speed, and batch sizes, and the results as shown in Fig. \ref{results}.

The results of Table. \ref{BPW} and Table. \ref{KLD} indicate that extended prompt table increases embedding capacity by 2.5 times and decreases KLD, while the change in anti-steganalysis ability is almost negligible. Therefore, we can both improve bpw and keep the effect of anti-steganalysis ability by improving the capacity of prompt table.

According to Fig. \ref{bpw}, batch size is positively correlated with bpw, and with the increase in batch sizes, bpw also keeps rising. Moreover, as shown in Fig. \ref{times}, as batch size increases, the time required decreases. Fig. \ref{bpw} and Fig. \ref{times} both  demonstrate that both bpw and time efficiency improve with increasing batch sizes in resource-rich situations.

\section{Conclusion}
In this paper, we propose a novel generative text steganography method based on diffusion model, named GTSD, which introduces the diffusion model into text steganography for the first time. This method improves generative speed, robustness, and imperceptibility while maintaining security. Specifically, we propose a new steganography scheme that embeds secret information through prompt mapping and batch mapping. Prompt mapping converts secret information into a conditional prompt to guide the pre-trained diffusion model in generating batches of candidate sentences. Batch mapping then selects the stego text based on the secret information from these batches. We conduct extensive experiments on the Movie, Twitter, and News datasets. The experimental results show that the proposed method outperforms the state-of-the-art method in terms of robustness, imperceptibility, and generative speed, while maintaining security. Furthermore, we demonstrate that the proposed GTSD has strong potential: as the prompt capacity and model batch sizes increase, the embedding capacity also increases, while maintaining security. In future work, we will also explore multi-language text steganography and improve the robustness of text steganography against other attacks.

\subsubsection{Acknowledgments.}
This work was supported by the National Natural Science Foundation of China (No. 62272463)

%
%
%

%

\begin{thebibliography}{10}
\providecommand{\url}[1]{\texttt{#1}}
\providecommand{\urlprefix}{URL }
\providecommand{\doi}[1]{https://doi.org/#1}

\bibitem{liao2020adaptive}
Liao, X., Yin, J., Chen, M., Qin, Z.: Adaptive payload distribution in multiple images steganography based on image texture features. IEEE Transactions on Dependable and Secure Computing  \textbf{19}(2),  897--911 (2020)

\bibitem{wu2020audio}
Wu, J., Chen, B., Luo, W., Fang, Y.: Audio steganography based on iterative adversarial attacks against convolutional neural networks. IEEE transactions on information forensics and security  \textbf{15},  2282--2294 (2020)

\bibitem{li2022anti}
Li, Z., Jiang, X., Dong, Y., Meng, L., Sun, T.: An anti-steganalysis hevc video steganography with high performance based on cnn and pu partition modes. IEEE Transactions on Dependable and Secure Computing  \textbf{20}(1),  606--619 (2022)

\bibitem{zhou2021linguistic}
Zhou, X., Peng, W., Yang, B., Wen, J., Xue, Y., Zhong, P.: Linguistic steganography based on adaptive probability distribution. IEEE Transactions on Dependable and Secure Computing  \textbf{19}(5),  2982--2997 (2021)

\bibitem{zhang2021provably}
Zhang, S., Yang, Z., Yang, J., Huang, Y.: Provably secure generative linguistic steganography. In: ACL (2021)

\bibitem{ding2023discop}
Ding, J., Chen, K., Wang, Y., Zhao, N., Zhang, W., Yu, N.: Discop: Provably secure steganography in practice based on" distribution copies". In: 2023 IEEE Symposium on Security and Privacy (SP). pp. 2238--2255. IEEE (2023)

\bibitem{yang2020vae}
Yang, Z.L., Zhang, S.Y., Hu, Y.T., Hu, Z.W., Huang, Y.F.: Vae-stega: linguistic steganography based on variational auto-encoder. IEEE Transactions on Information Forensics and Security  \textbf{16},  880--895 (2020)

\bibitem{mahato2013modified}
Mahato, S., Yadav, D.K., Khan, D.A.: A modified approach to text steganography using hypertext markup language. In: ACCT (2013)

\bibitem{ekodeck2016pdf}
Ekodeck, S.G.R., Ndoundam, R.: Pdf steganography based on chinese remainder theorem. Journal of information security and applications  \textbf{29},  1--15 (2016)

\bibitem{Chen2008LinguisticSD}
Chen, Z., Huang, L., Yu, Z., Yang, W., Li, L., Zheng, X., Zhao, X.: Linguistic steganography detection using statistical characteristics of correlations between words. In: Information Hiding (2008)

\bibitem{DBLP}
Xiang, L., Wang, X., Yang, C., Liu, P.: A novel linguistic steganography based on synonym run-length encoding. {IEICE} Trans. Inf. Syst.  \textbf{100-D}(2),  313--322 (2017)

\bibitem{wu2024ar}
Wu, T., Fan, Z., Liu, X., Zheng, H.T., Gong, Y., Jiao, J., Li, J., Guo, J., Duan, N., Chen, W., et~al.: Ar-diffusion: Auto-regressive diffusion model for text generation. Advances in Neural Information Processing Systems  \textbf{36} (2024)

\bibitem{van2003efficient}
Van~Le, T.: Efficient provably secure public key steganography. Cryptology ePrint Archive  (2003)

\bibitem{yang2018rnn}
Yang, Z.L., Guo, X.Q., Chen, Z.M., Huang, Y.F., Zhang, Y.J.: Rnn-stega: Linguistic steganography based on recurrent neural networks. IEEE Transactions on Information Forensics and Security  \textbf{14}(5),  1280--1295 (2018)

\bibitem{4370824}
Zhang, H.J., Tang, H.J.: A novel image steganography algorithm against statistical analysis. In: 2007 International Conference on Machine Learning and Cybernetics. pp. 3884--3888 (2007)

\bibitem{strudel2022self}
Strudel, R., Tallec, C., Altch{\'e}, F., Du, Y., Ganin, Y., Mensch, A., Grathwohl, W., Savinov, N., Dieleman, S., Sifre, L., et~al.: Self-conditioned embedding diffusion for text generation. ICLR  (2022)

\bibitem{li2022diffusion}
Li, X., Thickstun, J., Gulrajani, I., Liang, P.S., Hashimoto, T.B.: Diffusion-lm improves controllable text generation. Advances in Neural Information Processing Systems  \textbf{35},  4328--4343 (2022)

\bibitem{gong2022diffuseq}
Gong, S., Li, M., Feng, J., Wu, Z., Kong, L.: Diffuseq: Sequence to sequence text generation with diffusion models. In: ICLR (2022)

\bibitem{maas-etal-2011-learning}
Maas, A.L., Daly, R.E., Pham, P.T., Huang, D., Ng, A.Y., Potts, C.: Learning word vectors for sentiment analysis. In: ACL (2011)

\bibitem{go2009twitter}
Go, A., Bhayani, R., Huang, L.: Twitter sentiment classification using distant supervision. CS224N project report, Stanford  \textbf{1}(12), ~2009 (2009)

\bibitem{peng2021real}
Peng, W., Zhang, J., Xue, Y., Yang, Z.: Real-time text steganalysis based on multi-stage transfer learning. IEEE Signal Processing Letters  \textbf{28},  1510--1514 (2021)

\bibitem{wen2019convolutional}
Wen, J., Zhou, X., Zhong, P., Xue, Y.: Convolutional neural network based text steganalysis. IEEE Signal Processing Letters  \textbf{26}(3),  460--464 (2019)

\bibitem{yang2019ts}
Yang, Z., Wang, K., Li, J., Huang, Y., Zhang, Y.J.: Ts-rnn: Text steganalysis based on recurrent neural networks. IEEE Signal Processing Letters  \textbf{26}(12) (2019)

\bibitem{yang2020linguistic}
Yang, H., Bao, Y., Yang, Z., Liu, S., Huang, Y., Jiao, S.: Linguistic steganalysis via densely connected lstm with feature pyramid. In: Proceedings of the 2020 ACM Workshop on Information Hiding and Multimedia Security (2020)

\end{thebibliography}
\end{document}